\documentclass[preprint,12pt]{elsarticle}

\usepackage[margin=1in]{geometry}
\usepackage{amsmath}
\usepackage{graphicx}
\usepackage{inputenc}
\usepackage{amssymb}
\usepackage{natbib}
\usepackage{cancel}
\usepackage{caption}
\usepackage{lineno,hyperref}
\modulolinenumbers[5]
\usepackage[font=scriptsize]{caption}

\newcommand{\bea}{\begin{eqnarray}}  
\newcommand{\eea}{\end{eqnarray}} 

\journal{Physics Letters B}

\begin{document}

\begin{frontmatter}

\title{Model Building with Non-Compact Cosets}
\author{Djuna Lize Croon}
\ead{d.croon@sussex.ac.uk}
\address{University of Sussex, Brighton BN1 9RH}

\begin{abstract}
We explore Goldstone boson potentials in non-compact cosets of the form SO(n,1)/SO(n). We employ a geometric approach to find the scalar potential, and focus on the conditions under which it is compact in the large field limit. We show that such a potential is found for a specific misalignment of the vacuum. This result has applications in different contexts, such as in Composite Higgs scenarios and theories for the Early Universe. We work out an example of inflation based on a non-compact coset which makes predictions which are consistent with the current observational data.
\end{abstract}

\begin{keyword}
Model building \sep Cosmology \sep Non-compactness \sep Inflation 
\end{keyword}

\end{frontmatter}
\linenumbers

\section{Introduction}
Goldstone bosons are popular actors in theories beyond the Standard Model of Particle Physics. They resolve the dichotomy between the aptness of scalars in cosmological theories and the theoretical hierarchy problems that fundamental scalars suffer. The study of their Effective Field Theory is further motivated by their omnipresence in UV theories with global symmetries, such as models for axions \cite{Peccei:1977hh,Weinberg:1977ma,Wilczek:1977pj}, and supersymmetry \cite{Iliopoulos:1974zv,Fayet:1974jb,O'Raifeartaigh:1975pr}.

There is a vast body of literature which focuses on Goldstone bosons in compact cosets, that is, on theories in which a compact global symmetry breaks spontaneously to its compact subgroup. An example is the Minimal Composite Higgs Model MCHM$_5$, in which SO(5)$\rightarrow$SO(4). In theories of this kind the Goldstone bosons lie on a compact manifold, such as the hypersphere $S^4 \simeq SO(5)/SO(4)$. 
Their interactions are invariant under a shift symmetry, such that a potential is forbidden at all orders in perturbation theory. 

In the presence of a source of explicit breaking of the global group the shift symmetry is broken, and a potential for the pseudo-Goldstone Bosons (pGBs) may be generated. Such an explicit breaking can be mediated by external gauge bosons which gauge part of the global group, as is common in Composite Higgs models, or by couplings to instantons which do not respect the symmetry, as is the case with axions. 
The resulting potential will have a remnant periodic shift symmetry, stabilizing it against quantum corrections.
Examples which employ such a scenario are Composite Higgs models\cite{Agashe:2004rs}, Natural Inflation \cite{Freese:1990rb}, Goldstone Inflation \cite{Croon:2015fza,Croon:2015naa}, and composite dark matter \cite{Frigerio:2012uc}.

Goldstone bosons in non-compact cosets have received far less attention. Of particular interest are models in which a non-compact group breaks to its compact subgroup. There are indications that such cosets could give promising models of inflation \cite{Burgess:2014tja} and electroweak symmetry breaking \cite{Alonso:2016btr}. Like in the compact case, these cosets may address hierarchy problems by giving rise to stable scalar potentials. 

Here we will explore the idea that scalar sectors can be studied in a coordinate-invariant way, something that has recently attracted some attention in the context of Higgs Effective Field Theory \cite{Alonso:2015fsp,Alonso:2016btr,Alonso:2016oah}. It has been observed \cite{Low:2014oga} that results in non-compact cosets may be extrapolated from corresponding compact cosets by considering imaginary parameters, such that the corresponding manifold undergoes a Wick rotation. Here we instead follow a more general, geometric approach to study the potential of the Goldstone bosons of the hyperbolic space SO(n,1)/SO(n). In section \ref{Models} we describe the different models for hyperboloids that are of interest to this analysis. 

The shift symmetry in the non-compact case will also be broken in the presence of explicit symmetry breaking effects, misaligned with the original breaking. For SO(n,1)/SO(n) the remnant symmetry takes the form of a discrete scaling symmetry. We will parametrize the explicit breaking without choosing a particular particle physics interpretation, bearing in mind the different ways of breaking the shift symmetry.  
Our approach generalizes the analysis of \cite{Burgess:2014tja} in the context of inflation, and provides an alternative description of the discussion of Goldstone bosons in non-compact cosets in \cite{Alonso:2016oah}. 

The focus of this paper will be on the conditions under which the Goldstone boson potential is bounded, i.e. confined to lie in a specific region in the limit in which the field excursion of the scalars is large. This is of particular interest for inflationary model building, as in typical scenarios one has to explain the gap between the magnitude of the scalar potential ($V^{1/4} \sim 10^{15}$ GeV) and the large field excursion ($\Delta \phi \sim M_p$), highlighted by the familiar Lyth bound \cite{Lyth:1996im}. In section \ref{Potentials} we will show that a bounded potential is generated when the symmetry breaking parameters transform as a null vector of the hyperbolic space. 

In the last section we will discuss the application of this class of models to inflation. We will explore the inflationary predictions, and compare them to data from the Planck collaboration \cite{Ade:2015lrj}.

\section{Models of hyperbolic space}\label{Models}
Below the scale of the spontaneous breaking SO(n,1)$\rightarrow$SO(n), the relevant degrees of freedom are a set of Goldstone bosons which lie on the non-compact, n-dimensional hyperbolic sheet given by SO(n,1)/SO(n). In the absence of any additional sources of breaking, the Goldstone bosons respect a shift symmetry which forbids a scalar potential. They will obtain a potential when they couple to a source of explicit breaking.
This is for instance the case if a smaller group is gauged by external bosons such as in Composite Higgs models. This case is well studied; it has for instance recently been discussed in \cite{Alonso:2016btr} in the context of Higgs Effective Field Theory. 
Here we use a less restrictive approach, in which we focus on the transformation properties of the symmetry breaking parameters which couple to the Goldstone bosons.

The coset SO(n,1/SO(n) can be described as a sheet of a space-like hyperbola,\footnote{The terminology in this chapter is adopted often in analogy with space-time symmetries, however, the reader is assured that we consider internal symmetries only in this paper.} defined by the interval
\bea L &=& \{ (x_1,\text{...},x_{n+1}) ):\, x_{n+1}^2 - x_{n}^2 - x_{n-1}^2 - \text{...} - x_1^2 = \ell^2 \, \, \text{and} \,\, x_{n+1}>0\} \\
ds_L^2 &=& \eta_{\alpha \beta} dx^\alpha dx^\beta = dx_{n+1}^2 - \sum_{i=1}^n dx_i^2 \eea
where $x_{n+1} >0$. 
This space is associated with the Hermitian form or dot product with the signature $(n,1)$,
\bea  g_{\mu \nu} x^\mu y^\nu = x_\mu y^\mu  = - x_1 y_1 + x_2 y_2+ \text{...}+x_n y_n\eea
It has constant negative curvature, 
\bea \mathcal{R}_{fieldspace} = n (1- n) < 0.\eea

As we will see in the following, the model "L"  is not always the most transparent choice to describe the features of the Goldstone potential. An alternative choice is the Poincare disk model, which is defined by
\bea J &=& \{ (x_1,\text{...},x_{n+1}) ):\, x_1^2 + \text{...} + x_{n+1}^2 = \ell^2 \, \, \text{and} \,\, x_{n+1}>0\} \\
ds_J^2 &=&  \frac{ dx_{n+1}^2+ \sum_{i=1}^n dx_i^2 }{x_{n+1}^2} \eea
This model is related to "L" by a central projection from the point $(-\ell,0,\text{...},0)$, 
\bea L\rightarrow J, \hspace{2cm} (x_1,\text{...},x_n,x_{n+1}) \mapsto (x_1 \ell/x_{n+1},\text{...},\, x_n \ell /x_{n+1}, \, \ell^2/x_{n+1}) \eea

Another choice is the Poincare Half plane model, which reduces to the well known complex projective coordinates often employed in supersymmetry for $n=2$. The Half plane model is defined by
\bea H &=& \{ (1,x_2,\text{...},x_{n+1}) ): x_{n+1}>0\} \\
ds_H^2 &=& \frac{dx_{n+1}^2+ \sum_{i=2}^n dx_i^2 }{x_{n+1}^2}\eea
The Half plane model can in turn be related to J by a central projection from the point $(0,\text{...},0,\ell)$, i.e. the mapping
\bea J\rightarrow H, \hspace{2cm} (x_1,\text{...},x_n,x_{n+1}) \mapsto (-\ell, \, 2 \ell x_2/(x_1-\ell),\text{...},\, 2\ell  x_{n+1}/(x_1-\ell)) \eea
From this, it follows that "H" and "L" are related by the mapping, 
\bea L\rightarrow H, \hspace{2cm} (x_1,\text{...},x_n,x_{n+1}) \mapsto (-\ell, \, 2 \ell x_2/(x_1-x_{n+1}),\text{...},\, 2\ell^2/(x_1-x_{n+1})) \eea

\begin{figure}[h]
\centering 
  \includegraphics[width= 200pt]{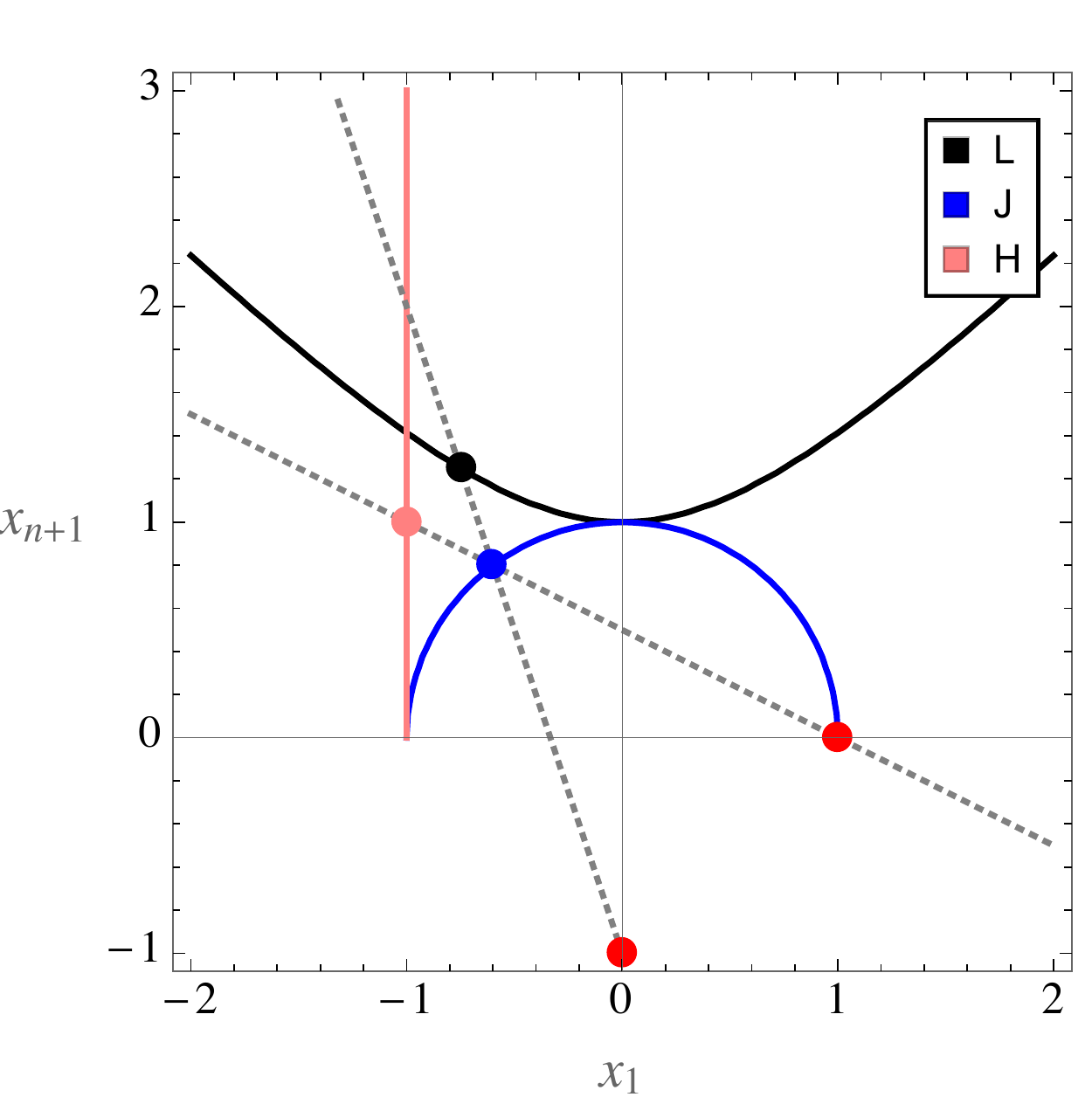}
  \caption{{\it Models of the coset:} in two dimensions. An arbitrary point on the hyperbola, $\{x^{^L}_1,x^{^L}_{n+1}\}$ (in black), can be projected it unto the sphere $x_1^2+x_{n+1}^2=1$ from the point $\{ -1,0\}$. This gives the coordinates in the Disk model, 
$\{ x_1^J,x_{n+1}^J \} = \left\{\frac{x^{^L}_1}{x^{^L}_{n+1}},\frac{1}{x^{^L}_{n+1}}\right\}$ (in blue). A further projection from the point $\{1,0\}$ onto the line $x_1=-1$ gives the coordinates in "H" 
$\{x_1^H,x_{n+1}^H \} = \left\{-1,2/\left(x_{n+1}^{^L}\left(1-\frac{x_1^{^L}}{x_{n+1}^{^L}}\right)\right)\right\}$ (in pink).}\label{para}
\end{figure}

\section{Compact potentials from non-compact cosets}\label{Potentials}
Before the second symmetry breaking, the Goldstone bosons are massless and their target metric is described by the hyperboloid SO(n,1)/SO(n). In the previous section we have shown different ways to describe such a field space. 

A potential for the Goldstone bosons of the coset SO(n,1)/SO(n) is generated in the presence of symmetry breaking effects, misaligned from the original vacuum. We use a very minimal description, based on particular choices for the transformation properties of the symmetry breaking parameters under the higher dimensional Lorentz group. Here we derive which transformation properties lead to a compact potential, i.e., a potential that does not diverge in the large field limit. 

First we observe the following,
 \bea x_{n+1}^{^L} - x_1^{^L} = 2\ell^2/x_{n+1}^H \eea
(and as we saw above $x_{n+1}^H>0$). This combination corresponds to an (n+1) dimensional null vector of SO(n,1). Notice that the symmetry always allows one to rotate the spacelike components of a vector into the $x_1$ direction. Thus if the symmetry breaking parameters transform as a null vector, that is,
\bea V_L =  \frac{V_\mu x^\mu}{\ell} = V  \left(\frac{x^L_{n+1} - x^L_1}{\ell} \right) \eea
where $V$ is just the normalization of the vector, we recover compactness in the ``H" coordinates,
\bea V_H =   \frac{2 \ell \, V}{x^H_{n+1}}.  \eea
The theory is then defined by this potential and the target metric in ``H", such that (dropping the superscripts)
\bea\label{Lkin} \mathcal{L} = \mathcal{L}_{kin} + V = \frac{f^2}{4} \frac{dx_2^2 + \text{...} + dx_{n+1}^2}{x_{n+1}^2}  +  \frac{V_1}{x_{n+1}} \eea
As this Lagrangian is invariant under a shift symmetry for all $x_k$  ($k \neq n+1$), it is seen that one can always find stationary points where $x_k$ is constant. In particular, it is always possible to find a field space trajectory for which only $x_{n+1}$ evolves. In that light, we may canonically normalize $x_{n+1}$ in terms of the field $\phi$,
\bea \label{redefphi} \phi = \frac{f}{\sqrt{2}} \log x_{n+1} \eea
such that we arrive at the negative exponential potential 
\bea \mathcal{L} = \frac{1}{2}(\partial \phi)^2 + V_1 e^{-\sqrt{2}\phi/f} \eea
Note that the positivity of $x_{n+1}$ guarantees that $\phi$ is a real direction and that $ - \infty < \phi \leq \infty$.

\begin{figure}[h] 
\centering
  \includegraphics[width= 200pt]{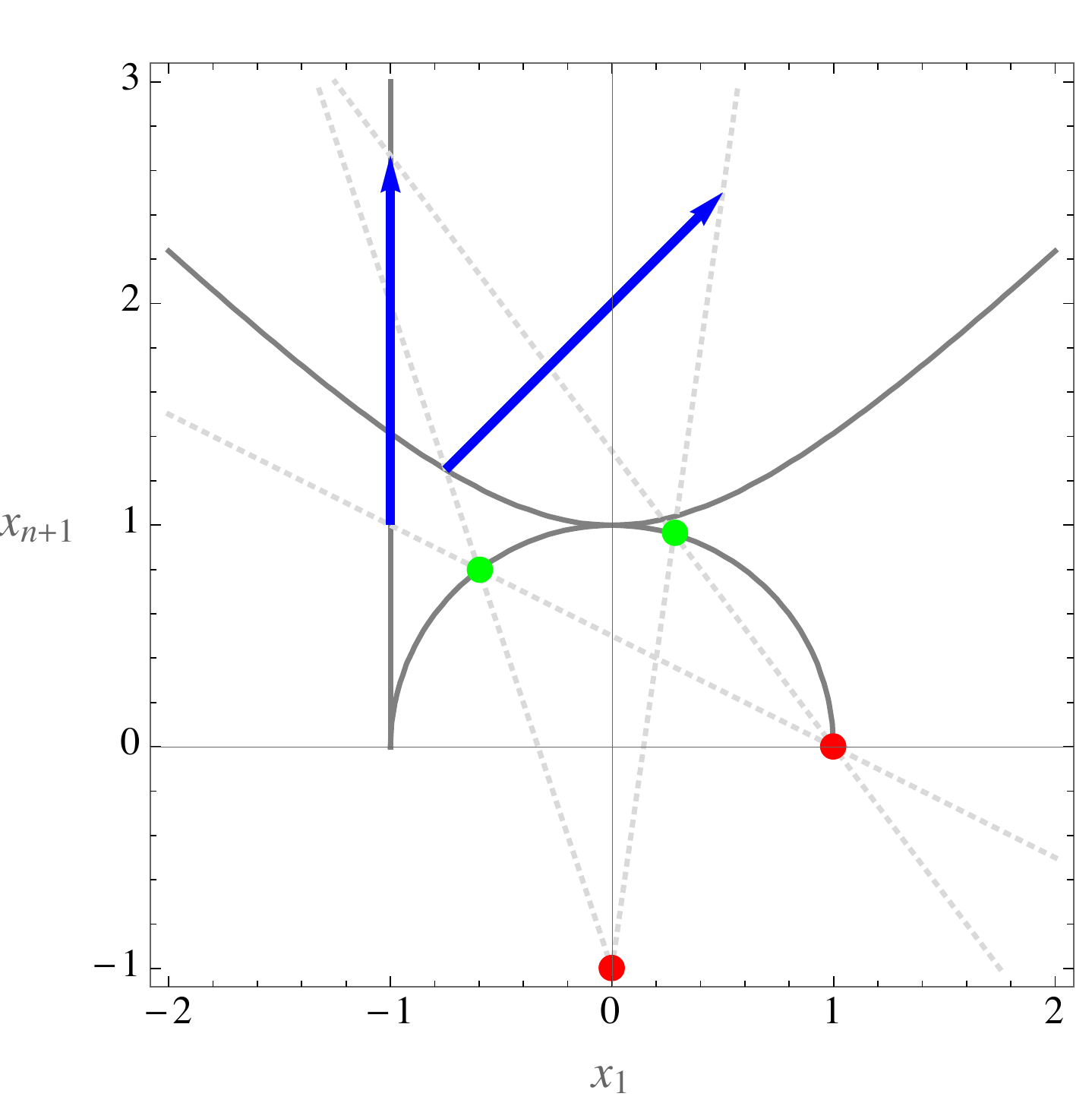}
  \caption{{\it Mapping of a transformation:} in two dimensions. It is seen that a time-like vector from an arbitrary point on the hyperbolic manifold will map to a finite projection in the model "H".}\label{para}
\end{figure}

Similarly, a timelike vector is mapped to its inverse by $L \rightarrow J$,
 \bea x_{n+1}^{^L} = \ell^2/x_{n+1}^J \eea
Following similar steps in this simpler example we arrive at the same result
\bea \mathcal{L} = \frac{1}{2} (\partial \phi)^2 + V_1 e^{-\sqrt{2}\phi/f} \eea
Note that this exponential potential respects a classical scale invariance, under which 
\bea \phi \rightarrow \phi + \epsilon,  \,\,\,\,\,\,\,\,\, x^\mu \rightarrow x^\mu e^{\epsilon/\sqrt{2}f}. \eea
and the action is just rescaled by $e^{\sqrt{2}\epsilon/f}$.

In general, the transformation rule of the symmetry breaking parameters allows for a constant term and higher order terms as well, suppressed by the relevant scale $\Lambda$. If the symmetry breaking effects can be parametrized in terms of a single vector $V_\mu$ which couples linearly to the Goldstones, higher order terms can be built from invariants of the form $V_\mu x^\mu$, such that, 
\bea V_L = V_0 + \sum_{j=1}^\infty g_j \, \frac{(\frac{V_\mu x^\mu}{\ell})^j}{\Lambda^{4(j-1)}} = V_0+  \sum_{j=1}^\infty   \tilde{g}_j \Lambda^4 \left(\frac{x_{n+1} - x_1}{\ell} \right)^j \eea
where $ \tilde{g}_j = g_j (V_1/\Lambda^4)^j$. 
The above steps will generate the exponential potential 
\bea  V &=& V_0 - V_1 e^{-\sqrt{2}\phi/f} +  \sum_{j=2}^\infty (-1)^j \,\tilde{g}_j \Lambda^4 e^{-j \sqrt{2}\phi/f}   \eea
which also respects the scaling symmetry, as it must. Generically one expects the higher harmonics to be of the size $(V_1/\Lambda^4)^j$ \cite{Dimopoulos:2005ac}. In this limit ($g_j=1$), the potential can be resummed, 
\bea V= V_0 + V_1  \sum_{j=1}^\infty \left(\frac{V_1}{\Lambda^{4}} \right)^{j-1} (-e^{- \sqrt{2}\phi/f})^j =V_0 -V_1\, \frac{ e^{- \sqrt{2}\phi/f}}{1+ (V_1/\Lambda^4) e^{- \sqrt{2}\phi/f}} \label{fullpot} \eea

We have seen that one may recover a bounded single field potential for any $n$, when the symmetry breaking dynamics transforms as a vector of SO(n,1). The dimensionality of the potential is unsurprising when one considers that an SO(n) transformation may always rotate the spacelike components along one direction, such that the only distinct cases are timelike, spacelike, and null. 

In the limit of exactly massless, non-interacting spectator fields, we do not expect the phenomenology to be altered with respect to the single field case. Let us consider the massless Goldstone fields in \eqref{Lkin}, with the field redefinition
\bea \label{redefchi} \chi_i = x_i f/\sqrt{2} \,\,\,\,\,\,\,\,\,\, i \neq n+1 \eea 
such that GBs have mass dimension 1. 
As only derivative interactions respect the shift symmetry for the (n-1) massless Goldstones, the lowest dimension couplings to fermions are dimension $d=5$ and of the form $\partial_\mu \chi_i \bar{\psi}\gamma^\mu \psi$.

Another possibility is that the (n-1) massless Goldstones are "eaten" to become the longitudinal components of gauge bosons, if the misalignment of the vacuum is due to gauging a subgroup of SO(n,1). Examples of such subgroups are given in Table \ref{Eating}. The effective mass of these gauge bosons will be set by the field $\phi$, which may develop a vacuum expectation value, reminiscent of the Higgs mechanism. 
This vev is dependent on the form of the scalar potential, which is sensitive to deviations from $g_j=1$, but is expected to be at least of the order of they symmetry breaking scale $f$.

\begin{table}[h]
\begin{center}
\begin{tabular}{ l | l  }
n&		Subgroup \\
\hline
9&		SU(3)\\
8&		SU(2)xSU(2)xU(1)\\
7&		SU(2)xSU(2) or SO(4)\\
6&		SU(2)xU(1)xU(1)\\
5&		SU(2)xU(1)\\
4&		SU(2) or SO(3)\\
3&		U(1)xU(1)\\
2&		U(1)\\
\end{tabular}
\end{center}
\caption{Gauged subgroups such that n-1 Goldstone Bosons are "eaten".}
\label{Eating}
\end{table}

\section{Inflation along the compact direction}\label{Inflation}
We have successfully constructed a compact potential in the large field limit based on the non-compact cosets SO(n,1)/SO(n). This is a promising candidate for an inflationary theory, in which one typically considers a large field excursion $\Delta \phi \sim M_p$, while measurements of the density perturbations forces the magnitude of the scalar potential to be orders of magnitude lower $V^{1/4}/\epsilon \sim 10^{15}$ GeV (where $\epsilon \ll 1$ is the first slow roll parameter). 

To interpret the pGB as the inflaton, consider the potential \eqref{fullpot},
\bea V = V_0 + V_1  \sum_{j=1}^\infty g_j \left(\frac{V_1}{\Lambda^{4}} \right)^{j-1} (-e^{- \sqrt{2}\phi/f})^j \eea
where here we phenomenologically impose $V_0=\Lambda^4$ to render the potential positive definite. 
We will work out an example with $g_j=1$ such that the potential is bounded from below, however, as we will find below, the inflationary results only depend strongly on the first term in the sum and are therefore insensitive to this assumption. 
With these assumptions, the potential can be rewritten as
\bea V= \frac{\Lambda^4}{\alpha  e^{-\sqrt{2}  \phi/f }+1} \eea
where $\alpha = V_1 / \Lambda^4$.
This model can reproduce the inflationary predictions from the Planck data \cite{Ade:2015lrj}. It is, to a first approximation, a single field model, with slow roll parameters given by 
\bea \epsilon = \frac{M_p^2}{2} \left(\frac{V'(\phi)}{V(\phi)}\right)^2 =\frac{M_p^2}{f^2} \frac{\alpha ^2}{  \left(\alpha +e^{\sqrt{2}\phi /f}\right)^2} \eea
\bea \eta  = M_p^2 \left(\frac{V''(\phi)}{V(\phi)}\right) - \frac{M_p^2}{2} \left(\frac{V'(\phi)}{V(\phi)}\right)^2 =  \frac{M_p^2}{f^2} \frac{\alpha  \left(\alpha -2 e^{\sqrt{2}\phi /f}\right)}{ \left(\alpha +e^{\sqrt{2}\phi /f}\right)^2} \eea
where $M_p$ is the reduced Planck mass. 
It is seen that for a slow roll scenario to take place below the Planck scale ($f < M_p$), one has to be in the regime $\alpha \ll e^{\sqrt{2} \phi /f}$. Thus, to study inflation it would have been sufficient to only consider the first term in the series expansion; we have considered the resummed potential here to show that it is bounded from below. 

Slow roll ends when $\epsilon \rightarrow 1$. We use this relation to estimate  
\bea \phi_E = \frac{\log (\alpha  (\beta -1))}{\sqrt{2} \beta } \eea
where $\beta = (f/M_p)^{-1}$.
The number of e-foldings is then given by 
\bea N = \frac{1}{M_p^2} \int_{\phi _i}^{\phi _e} \frac{  V(\phi )}{V'(\phi )} \, d\phi \eea

\begin{figure}[h] 
\centering
  \includegraphics[width= 300pt]{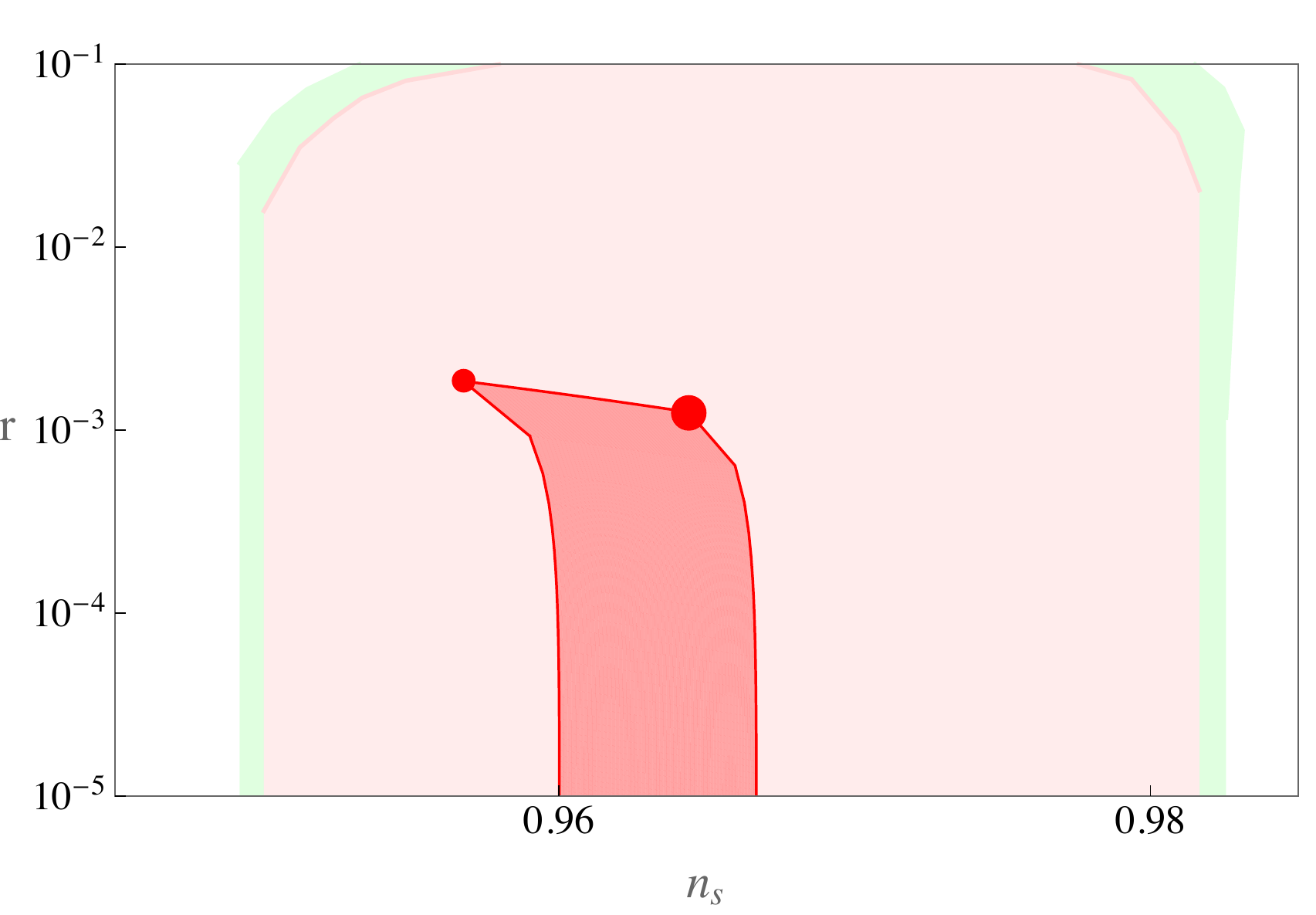}
  \caption{{\it Inflationary predictions: $n_s$ versus $r$.} The biggest dot marks the point $N=60$, $f \approx M_p$, the smaller dot marks $N=50$, $f\approx M_p$. Lowering the scale $f$ corresponds to lowering the tensor to scalar ratio, as indicated by the sweeping lines in the plot. In green the TT spectrum and polarisation data at low-$\ell$ (lowP) from \cite{Ade:2015lrj}; in pink the combined spectra TT, TE, EE +lowP.}\label{nsr}
\end{figure}

In figure \ref{nsr} we show the inflationary predictions of the model in terms of the spectral index $n_s$ and the tensor- to scalar ratio $r$. The values that fall are allowed by the Planck data, and satisfy 
\bea \label{ftoMp} \frac{f}{M_p} < 1 \eea
lowering this ratio implies reducing the prediction for the tensor to scalar ratio, as is common to Goldstone Inflation models \cite{Croon:2015fza,Croon:2015naa}.

As shown in the previous section, the other GB fields ($\chi_i = x_i f/\sqrt{2}$, $i \neq n+1$ ) are exactly massless in this model, and the inflationary dynamics is therefore completely dominated by the inflaton field $\phi$. 
It is well known that for light spectator fields with $V''(\chi_i)/H^2 \ll 1$ the non-Gaussianity parameter $f_{NL}$ is suppressed and there is no observable effect on the inflationary power spectrum. This effect can be checked explicitly for the current model (with a non-trivial field space metric) using the methodology described in \cite{astro-ph/0701247}.

However, the effects of the massless GB fields may become manifest during reheating after the slow-roll phase. 
The leading couplings between the massive field $\phi$ and other fields carry at least dimension five. 
A two-body decay to Standard Model gauge bosons from a coupling to the field strength would for instance be given by the decay rate $\Gamma_{\phi F^2} = 1/(8 \pi) |\mathcal{M}|^2/m_\phi \simeq 1/(8 \pi)  m_\phi^3/M_p^2$.
When $\phi$ is interpreted as the inflaton, this may lower the reheat temperature, typically given by $T_r= \sqrt{\Gamma_{\phi F^{2}}  M_p}$.

The coupled kinetic term \eqref{Lkin} mixes the evolution of the scalar fields.  As we will see, the curved field-space metric may give rise to a backreaction on the inflaton dynamics that forms the background for reheating. This effect is distinct from the usual Hubble induced mass for spectator fields (for instance such as described in \cite{1607.07058}).
From the equations \eqref{Lkin}, \eqref{redefphi}, \eqref{redefchi}, the equation of motion for the $\phi$ and $\chi_i$ fields are given by (conform \cite{Amin:2014eta})
 \bea \label{eomchi}  \ddot{\chi_i}  + 3 H  \dot{\chi_i} - \frac{2 \sqrt{2}}{f}  \dot{\chi_i}  \dot\phi &=& 0 \\  \label{eomphi} \ddot{\phi}  + 3 H \dot{\phi} + \frac{2 \sqrt{2}}{f} e^{- \frac{2 \sqrt{2} \phi  }{f}} ( \dot{\chi_i})^2 +  \frac{ \sqrt{2} V_1}{f} e^{-\frac{ \sqrt{2} \phi  }{f}}&=& 0 \eea
 Here the Hubble parameter is given by 
 \bea \label{Hubble} H^2 = \frac{1}{3 M_p^2} \left[ \frac{1}{2} \left(\dot\phi^2 + e^{- \frac{2 \sqrt{2} \phi  }{f}} \dot\chi_i^2 \right) + V(\phi)  \right] \eea
 where summation over the field index $i$ is implied in both \eqref{eomphi} and  \eqref{Hubble}. 
 The curved metric becomes important in the regime $ H  \ll \dot\phi/f$, for which the last term in \eqref{eomchi} dominates. 
In this limit the equation is solved by 
  \bea  \dot{\chi_i}  &=&  c_1 f^2 e^{\frac{2 \sqrt{2} \phi  }{f}} \eea
Inserting this result to solve the equation of motion for the inflaton, we find
\bea   \ddot{\phi} + 3 H \dot{\phi} + 2 \sqrt{2} \, (n-1) f^3  e^{ \frac{2 \sqrt{2} \phi  }{f}} +  \frac{ \sqrt{2} V_1}{f} e^{-\frac{ \sqrt{2} \phi  }{f}} &=& 0 \eea 
For instance, in the limit $\phi/f \ll 1$, this is solved by
\bea \phi /f \approx c_1 e^{i \tilde{f}_n^2 \,\frac{t }{f}}+c_2 e^{- i  \tilde{f}_n^2 \frac{ t}{f}}- \sqrt{2}  \frac{ \hat{f}^4_n}{ \tilde{f}_n^4 } \eea
where we have defined \bea \tilde{f}_n^4 &=&  2 \,(n-1) f^4 - V_1/2 \\ \hat{f}^4_n &=&  2\, (n-1) f^4 +V_1 .\eea
The coefficients $c_1$ and $c_2$ are set by the value of the field when inflation ends, usually estimated as $\phi/f |_{\epsilon=1} = c_1 + c_2 - \sqrt{2}$. For $V_1  \ll f^4$ the oscillation simplifies to 
\bea \phi /f \approx c_1 e^{i \sqrt{2 \,(n-1)} \,  f t}+c_2 e^{- i \sqrt{2 \,(n-1)} \,  f t}- \sqrt{2}   \eea
Thus we can see that the dimension of the symmetry breaking can play a role in the oscillation period of the inflaton after inflation, and thus affect timescale of reheating.

\section{Conclusions}
Here we have considered a geometric approach to describe Goldstone bosons in the non-compact coset SO(n,1)/SO(n), by considering different models to describe the hyperbolic field space. Since the dynamics of the Goldstone bosons is specified by the target metric of this manifold, this approach lends itself to simple visualizations. The Goldstone bosons obtain a potential when their shift symmetry is broken by terms that explicitely break SO(n,1) and couple to the Goldstone bosons. This can also be described geometrically, and without choosing a specific model, in terms of symmetry breaking parameters.

It was well known that pGB potentials based on compact cosets are compact, that is, the resulting potential is bounded in the large field limit. Here we have shown for cosets describing hyperbolic manifolds (in arbitrary dimensions) that the pGB potential can be compact as well. We have shown two examples, in which the parameters that break the global symmetry transform like a time-like or null vector of the hyperbolic space. This result is promising for models of inflation, and gives predictions compatible with the data from Planck \cite{Ade:2015lrj} as shown in Fig.\ref{nsr}.

Our result can be generalized by considering the breaking of different non compact cosets. For example, one could consider the breaking of an indefinite orthogonal group to its maximal (compact) subgroup, i.e. the coset $SO(p,q)/SO(p)$. In this case one may also use projective coordinates, to find that there exists a parameterization in which the vector $ (x_{i \in p}-x_{j \in q}) $ maps to $2 \ell / x_{j \in q}$. 

The geometric approach may also be applied in other contexts. The negative exponential potential is reminiscent of the first ekpyrotic models \cite{Lehners:2007ac}, though this case which has one effective degree of freedom does not reproduce a (nearly) scale invariant spectrum of perturbations. It remains an open question whether a similar result can be found for an effectively multifield model. 

The techniques described here may also find an application in Higgs physics, complimentary to recent HEFT studies in spaces with negative curvature \cite{Alonso:2015fsp,Alonso:2016btr,Alonso:2016oah}. Such a study may allow a unified understanding of the dynamics in an arbitrary number of dimensions. 

 \section*{Acknowledgement}
 DL would like to thank C. Burgess for useful discussions and the Perimeter Institute for the support and hospitality during the development of this work. 

\section*{References}
\bibliography{NonCompact}

\end{document}